\documentclass{article}

\usepackage{arxiv}

\usepackage[utf8]{inputenc} 
\usepackage[T1]{fontenc}    
\usepackage{hyperref}       
\usepackage{url}            
\usepackage{booktabs}       
\usepackage{amsfonts}       
\usepackage{amsmath} 
\usepackage{nicefrac}       
\usepackage{microtype}      
\usepackage{lipsum}
\usepackage{graphicx}
\usepackage{float}
\usepackage{natbib} 
\usepackage{comment}
\usepackage{enumitem}

\title{Non-Stationarity in Brain-Computer Interfaces: An Analytical Perspective}

\author{
 Hubert Cecotti \\
  Department of Computer Science\\
  California State University, Fresno\\
  Fresno, USA\\
   \And
Rashmi Mrugank Shah \\
  Department of Computer Science\\
  California State University, Fresno\\
  Fresno, USA\\
  \And
Raksha Jagadish \\
  Department of Computer Science\\
  California State University, Fresno\\
  Fresno, USA\\
 \And
Toshihisa Tanaka \\
  Department of Electrical Engineering and Computer Science\\
  Tokyo University of Agriculture and Technology\\
  Tokyo, Japan \\
}

\begin{document}
\maketitle

\begin{abstract}Non-invasive Brain-Computer Interface (BCI) systems based on electroencephalography (EEG) signals suffer from multiple obstacles to reach a wide adoption in clinical settings for communication or rehabilitation. Among these challenges, the non-stationarity of the EEG signal is a key problem as it leads to various changes in the signal. There are changes within a session, across sessions, and across individuals. Variations over time for a given individual must be carefully managed to improve the BCI performance, including its accuracy, reliability, and robustness over time. This review paper presents and discusses the causes of non-stationarity in the EEG signal, along with its consequences for BCI applications, including covariate shift. The paper reviews recent studies on covariate shift, focusing on methods for detecting and correcting this phenomenon. Signal processing and machine learning techniques can be employed to normalize the EEG signal and address the covariate shift.
\end{abstract}


\section{Introduction}
\label{section:introduction}

Brain–Computer Interfaces (BCIs) provide a direct communication link between neural activity and external devices, enabling interaction without relying on muscle movement~\cite{Krusienski_2011,Hinterberger2005,5626537}. Envisioned initially as assistive technologies for individuals with severe motor impairments, BCIs are now finding applications in diverse areas, including gaming, neurofeedback, cognitive workload monitoring, and human–machine collaboration~\cite{doi:10.26599/BSA.2020.9050017,6169943,6046114}. Among the various recording modalities explored, non-invasive electroencephalography (EEG) remains the most widely used because it is user-friendly, relatively inexpensive, portable, and offers excellent temporal resolution~\cite{6401177,10646518}. EEG-based BCIs typically detect specific neural signatures such as Motor Imagery (MI)~\cite{INCE2009236,ROY2022105347,8897723,6506618}, Steady-State Visual Evoked Potentials (SSVEP)~\cite{9516951}, and Event-Related Potentials (ERP) like the P300 component~\cite{6401177}. ERP-based paradigms, such as the P300 speller~\cite{farwell1988talking} or Rapid Serial Visual Presentation (RSVP)~\cite{luck1990electrophysiological}, have shown considerable promise in target detection and attention monitoring, i.e., passive BCI~\cite{Mousavi02102019}. Hybrid approaches that combine MI with SSVEP or ERP can enhance classification performance and increase system flexibility~\cite{7782750}.  

Despite these advances, EEG-based BCIs face a fundamental and persistent challenge: non-stationarity, the fact that the statistical properties of EEG signals change over time, across sessions, and between individuals~\cite{Krusienski_2011,8695803}. A machine learning model trained on one session or subject often performs poorly when applied to new data under otherwise identical conditions~\cite{Cho_2015,9441076,9175985}. Performance can also deteriorate within a session. This performance drop is commonly caused by \emph{covariate shift}, where the distribution of the input features changes while the relationship between features and labels remains the same~\cite{sugiyama2007covariate,5415628,10887776}. Such shifts can strongly influence key stages of BCI processing, including spatial filtering~\cite{7296710}. Contributing factors include user fatigue, changes in cognitive state, electrode displacement, and environmental noise~\cite{7280742,Krusienski_2011,doi:10.26599/BSA.2020.9050017}. In practical settings, these issues often make frequent recalibration unavoidable, limiting the scalability and long-term usability of BCIs~\cite{7288989,app132312800}.

Recently, more targeted strategies for addressing covariate shift have emerged, such as selecting specific EEG channels and trials that are most invariant across sessions~\cite{8914414,khalil2022novel}. Classifier performance under covariate shift has also been systematically reviewed and compared in various domains, including EEG-based movement classification~\cite{6854726}, acoustic emotion recognition~\cite{6488742}, and across species in Traumatic Brain Injury (TBI) research~\cite{vishwanath2023reducingintraspeciesinterspeciescovariate}. These advances underscore the need for principled approaches to covariate shift mitigation that are both data-efficient and generalizable~\cite{8978471}.

In the context of TBI research, covariate shift can arise when translating findings from animal models to humans. Rodent and human neural or neuroimaging datasets often differ in their statistical distributions due to physiological variations, differences in injury mechanisms, and disparities in recording setups, despite the assumption that the mapping between biomarkers and clinical outcomes is similar. Addressing these cross-species distributional differences is critical for the reliable translation of preclinical findings into effective human therapies. This example illustrates that covariate shift is not unique to EEG-based BCIs, but is a broader challenge across neuroscience and biomedical research~\cite{vishwanath2023reducingintraspeciesinterspeciescovariate}.


\subsection{Rationale and Contribution of This Review}
\label{subsec:rationale}

Although many techniques have been proposed to counteract EEG non-stationarity, the literature is fragmented. Some reviews focus narrowly on signal processing strategies, while others emphasize machine learning methods, leaving the broader picture underdeveloped~\cite{9134411}. It is often unclear whether the problems of non-stationarity that are addressed with machine learning could have been handled by signal processing methods. Efficient EEG signal normalization can reduce the impact of artefacts and temporal changes on the subsequent classification or detection stages. Important aspects, such as the role of latent neural dynamics or inter-subject similarities in shaping transfer learning performance, are rarely integrated into these discussions~\cite{10340490}. Systematic analysis of how datasets and evaluation protocols influence generalization remains limited ~\cite{Luo2024CrossSubject}.

This review aims to address these gaps. We present a structured, critical synthesis of recent work, organized around four guiding questions: (1) Why is mitigating non-stationarity crucial for real-world BCI applications? (2) What are the most effective techniques for doing so? (3) Which challenges remain unresolved? and (4) Which datasets are best suited for studying generalization in the face of non-stationarity?


\subsection{Why is Dealing with Non-Stationarity Critical for BCI Applications?}
\label{subsec:importance}

A BCI’s usefulness depends on its ability to operate reliably across different sessions, users, and environments. Non-stationarity undermines this by introducing variability into the signal, even when the task and recording setup are unchanged~\cite{sussillo2016making}. Aligning neural dynamics between calibration and deployment phases helps maintain performance stability in closed-loop control systems ~\cite{7782750}. Without such strategies, recalibration becomes unavoidable, consuming valuable user time and reducing system accessibility~\cite{Chang2024Multistrategy}. Determining the optimal timing for recalibration is essential when shifts between training and test data can significantly degrade performance ~\cite{Chang2024Multistrategy}.

This challenge is particularly acute in paradigms like RSVP, where moment-to-moment changes in attention, visual processing, and fatigue are common~\cite{Mousavi02102019}. Managing non-stationarity is therefore a prerequisite for designing adaptive, user-friendly BCIs suitable for everyday use~\cite{7782750,6346563}.

\subsection{Main Techniques for Handling Non-Stationarity}
\label{sec:techniques}

A variety of strategies have been developed to address non-stationarity, drawing from statistical learning, domain adaptation, and online model adaptation~\cite{7296710,6425489}. Some methods align training and test distributions before classification, while others update the model incrementally with new data ~\cite{10.1145/3615592.3616853,8013808}. Change-point detection applied to EEG covariance matrices can identify and correct abrupt distribution shifts during operation ~\cite{10887776}.
One foundational approach is \textbf{sample reweighting}, which adjusts the contribution of each training sample to match the target distribution better~\cite{sugiyama2007covariate}. Domain-invariant representation learning extracts features that remain stable across sessions or subjects ~\cite{9057712,9154600,ZHANG2021106150}. 
Adversarial domain adaptation aligns source and target feature distributions without requiring target labels~\cite{Farshchian2018Adversarial,9534286,9177281}. \textbf{Multi-source learning} combines data from multiple subjects or sessions to create more robust models~\cite{WANG2024106742,RAZA2015659,9057712}. 
Other promising directions include \textbf{online and adaptive learning}, where the model is updated continuously with streaming EEG data~\cite{4352845,6177271}, and \textbf{multi-session/subject training}, which increases model robustness by training on diverse data sources~\cite{DONG2023104453,ROY2022105347}. Optimal transport methods can accelerate adaptation by aligning source and target domains in a principled way ~\cite{10.1145/3615592.3616853}. Together, these techniques aim to reduce the need for frequent recalibration and improve long-term usability~\cite{7319297,Mousavi02102019,app132312800}.

\subsection{Current Challenges}
\label{sec:challenges}

Despite progress, several obstacles remain. \textbf{Limited generalization} continues to be a major issue, as even state-of-the-art algorithms struggle with substantial variability across subjects~\cite{7280742}. Cross-subject learning is particularly challenging because inter-individual neural differences often exceed task-related differences~\cite{Luo2024CrossSubject}.

Data scarcity compounds the problem: collecting large, labeled EEG datasets spanning multiple sessions and participants is time-intensive and expensive~\cite{Luo2024CrossSubject}. High computational demands of many algorithms further limit their real-time applicability ~\cite{4462222,Chang2024Multistrategy}.

A further complication is the lack of \textbf{standardized evaluation protocols} for non-stationarity, which hampers fair comparisons between methods~\cite{app132312800}. Finally, factors such as fluctuating attention, emotional state, fatigue, and hardware drift add unpredictable variability~\cite{Shen2019_challenge_aBCI,MILADINOVIC2021105808}, underscoring the complexity of building reliable, calibration-free BCIs~\cite{Krusienski_2011}.

\subsection{Datasets for Studying EEG Non-Stationarity}
\label{sec:datasets}

A number of publicly available datasets are valuable for investigating non-stationarity and generalization. BNCI Horizon 2020 datasets (e.g., 001-2014, 004-2015) provide multi-session motor imagery recordings from multiple users ~\cite{ZHANG2021106150}. OpenBMI and Korea University datasets contain both MI and RSVP paradigms recorded over several sessions, supporting cross-session adaptation research ~\cite{9195541}. These datasets are frequently used to evaluate Riemannian geometry-based and covariance alignment methods ~\cite{7296710}.

The EPFL-BCI dataset includes RSVP-based EEG recordings from multiple participants, enabling research on attention-related ERPs ~\cite{6401177}. The UAM P300 dataset provides long-term, session-spanning P300 data for studying ERP classifier stability ~\cite{4785192}. Large, heterogeneous datasets are also important for benchmarking adversarial and optimal transport-based transfer learning approaches ~\cite{10.1145/3615592.3616853}. Together, these resources enable systematic evaluation of algorithms designed to handle covariate shift, session drift, and subject variability~\cite{7288989}.

\subsection{Contributions and Outline}
\label{sec:outline}

This paper makes several contributions: (1) a focused overview of the non-stationarity problem in EEG-based BCIs and its implications for real-world deployment, especially in RSVP paradigms; (2) a survey and categorization of strategies for mitigating non-stationarity, including domain adaptation, transfer learning, and online learning techniques; (3) identification of persistent challenges hindering robust and calibration-free BCIs; and (4) a review of widely used EEG datasets relevant to non-stationarity and generalization.

The paper is organized into four main sections: Section \ref{section: Non-stationary} discusses the causes, implications, and challenges of EEG non-stationarity; Section \ref{section:review} describes the review methodology; Section \ref{section: approaches} presents a structured overview of approaches for handling non-stationarity, including signal normalization, feature alignment, classifier adaptation, and domain adaptation strategies, along with recent advances in reinforcement learning–based and hybrid methods; Section \ref{section:discussion} synthesizes findings, evaluates existing methods, and highlights research gaps; and Section \ref{section:conclusion} summarizes the main insights and emphasizes the importance of addressing non-stationarity in EEG for robust BCI systems.

\subsection{Summary}
\label{sec:summary}

The effectiveness of EEG-based BCIs—particularly RSVP-based systems—depends on managing non-stationarity and covariate shift in neural signals. This introduction has addressed four central questions: the critical role of non-stationarity in limiting generalization and usability; the range of mitigation techniques, including domain adaptation and adaptive learning; the technical and practical challenges that persist; and the availability of benchmark datasets for studying non-stationarity in real-world BCI contexts.

These discussions establish the foundation for the remainder of the paper, which provides a detailed review of methods, compares their effectiveness, and outlines future directions for building more robust, adaptive, and user-friendly BCI systems.

\section{Non-stationarity in the EEG signal}
\label{section: Non-stationary}

EEG signals form the backbone of most non-invasive BCI systems. While EEG offers a safe and accessible means of measuring brain activity, its use in practical BCI applications is hindered by the inherent non-stationarity of the signals. In signal processing terms, a time series is said to be non-stationary if its statistical properties—such as mean, variance, autocorrelation, and frequency spectrum—change over time. EEG signals exhibit such variability not only across sessions and individuals but also within a single recording session, even when the subject performs the same task under seemingly identical conditions~\cite{7280742,Cho_2015,PHILIP2022824}. This section provides an in-depth analysis of the sources, types, and consequences of EEG signal non-stationarity in the context of BCI research and development.

\subsection{Understanding Non-Stationarity in EEG-based BCIs}

The deployment of machine learning models in BCI systems assumes that training and test data are drawn from the same underlying distribution. However, EEG signal distributions often violate this assumption due to non-stationarity, resulting in degraded classifier performance, increased calibration needs, and reduced user trust~\cite{9441076,RAZA2015659,8852284}. As a result, BCI systems frequently suffer from poor generalization across sessions or subjects unless explicitly designed to adapt to or compensate for these distributional shifts.

This variability is closely linked to the phenomenon of covariate shift, where the marginal distribution of inputs $P(X)$ changes between training and deployment, while the conditional distribution $P(Y|X)$ remains constant ~\cite{5415628,sugiyama2007covariate,6999160}. In EEG-based BCIs, covariate shift can manifest when the same cognitive task (e.g., detecting a P300 in RSVP paradigms) elicits differing EEG patterns across users or sessions. Therefore, the non-stationary nature of EEG signals underpins both technical and practical challenges in building reliable BCI systems.

EEG non-stationarity can be categorized into three distinct levels~\cite{raza2016adaptivethesis}:

\begin{enumerate}[nosep]
    \item \textbf{Within-session variability}: Short-term changes during a single session due to fatigue, attention fluctuation, or emotional state.
    \item \textbf{Between-session variability}: Day-to-day differences due to electrode shift, physiological changes, learning effects, or external conditions.
    \item \textbf{Inter-subject variability}: Structural and functional differences in the brains of different users.
\end{enumerate}

This structured view facilitates the design of mitigation strategies tailored to the specific type of non-stationarity encountered.

\subsection{Causes of EEG Signal Non-Stationarity}

A comprehensive understanding of the underlying causes is essential for tackling EEG non-stationarity. Based on an extensive review of existing literature, including recent developments in both EEG and MEG data analysis~\cite{PHILIP2022824}, the main contributing factors are detailed below.

\subsubsection{Physiological and Cognitive Fluctuations}

\begin{figure}[H]
  \centering
  \includegraphics[width=1\columnwidth]{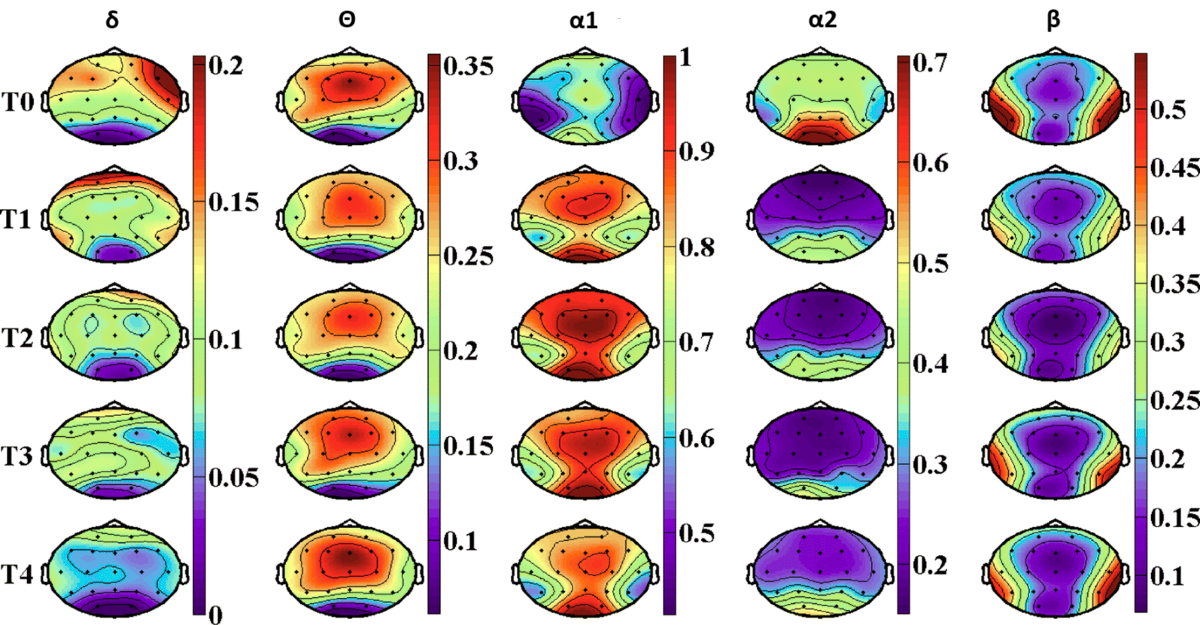}
  \caption{EEG relative power topography in theta, alpha, and beta bands showing variability due to cognitive and physiological fluctuations, such as fatigue and attention. Adapted from Lin et al.~\cite{Lin2020_fatigue_topography}.}
  \label{fig:physio_fluctuations}
\end{figure}

One of the primary sources of non-stationarity stems from internal physiological and cognitive changes. EEG signals are susceptible to fluctuations in attention, alertness, mental fatigue, mood, motivation, and stress. These internal states modulate cortical rhythms, leading to variations in alpha, beta, and theta band activity over time~\cite{8512974, Wan2023_nonstationarity_spikes, 4100850}. Fatigue is associated with a decreased signal-to-noise ratio and increased theta activity, which can hinder the accurate extraction and classification of features. Emotional states such as anxiety or frustration, especially in feedback-based BCI systems, can further introduce variability by altering autonomic and neural responses~\cite{KAPGATE2024100109}.

As shown in Figure~\ref{fig:physio_fluctuations}, these fluctuations can cause moment-to-moment shifts in signal amplitude and baseline drift, simulating real-world non-stationarity. Peripheral physiological signals—like heart rate, respiration, and galvanic skin response—interact with cortical activity and contribute to EEG variability, even in passive paradigms~\cite{6609172}.

\subsubsection{Inter- and Intra-Subject Variability}

Variability exists both across users and within the same user over time. \textbf{Inter-subject variability} is influenced by anatomical differences such as skull thickness, cortical folding, and individual neural efficiency. These differences result in unique signal topographies that often require user-specific model calibration~\cite{s21093225, 6945117}.

\begin{figure}[H]
    \centering
    \includegraphics[width=0.9\linewidth]{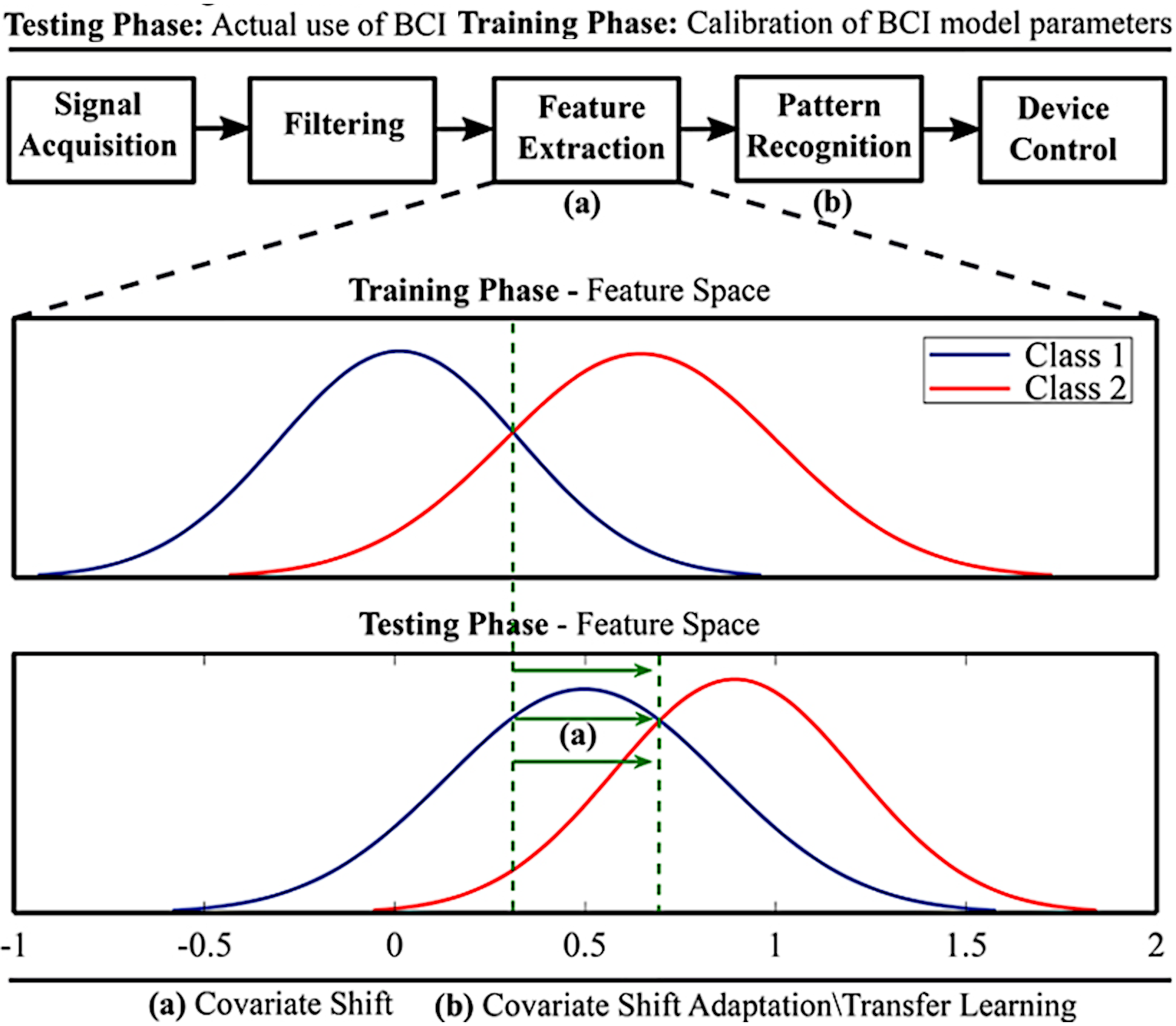}
    \caption{Illustration adapted from Saha and Baumert (2019)~\citep{saha2019intra}, showing intra- and inter-subject variability in EEG-based sensorimotor BCI. Inter-subject variability (top) reflects differences in EEG feature distributions across individuals due to anatomical and functional differences, while intra-subject variability (bottom) captures session-to-session changes within the same individual caused by physiological or psychological fluctuations. Source: Adapted from \textit{Frontiers in Computational Neuroscience}. ~\citep{saha2019intra}}

    \label{fig:inter_intra_variability}
\end{figure}

\textbf{Intra-subject variability}, on the other hand, encompasses changes in a user's EEG signals across different sessions or even within a single session. Factors such as hydration, sleep quality, stress levels, and hormonal cycles can alter brain dynamics, leading to inconsistent signal patterns~\cite{app132312800,7319297,MILADINOVIC2021105808}. 

Figure~\ref{fig:inter_intra_variability} illustrates these two forms of variability, both of which hinder the development of universal, user-independent BCI models.

\subsubsection{Electrode Shift and Impedance Variations}

Even small changes in the position of electrodes on the scalp can lead to significant differences in recorded EEG signals, as the spatial projection of neural activity onto the scalp is highly sensitive to electrode location~\cite{app132312800}. Variations in contact impedance—caused by gel drying, skin perspiration, cap tension, or electrode pressure—introduce broadband noise and amplitude distortions ~\cite{app132312800, 4811472,Samek_2012,6945117}.

\begin{figure}[H]
    \centering
    \includegraphics[width=1.1\textwidth]{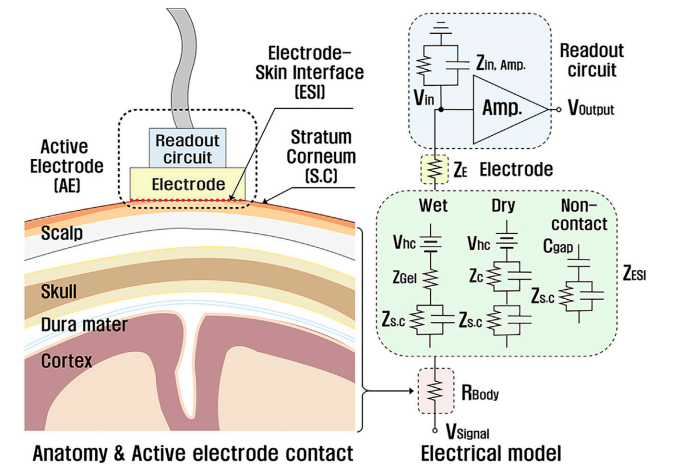}
   \caption{An anatomical graphic of the EEG signal path and the corresponding electrical model, illustrating electrode-skin interface, conductive gel, and tissue layers. Variations in these impedances influence EEG signal quality and contribute to noise and distortion, which are key factors affecting EEG non-stationarity and classification robustness. Adapted from Seok et al. (2021) \citep{Seok2021}.}
    \label{fig:eeg_impedance_model}
\end{figure}

Figure~\ref{fig:eeg_impedance_model} shows the anatomical and electrical impedance model of the EEG signal path. These hardware-related factors are a key source of between-session non-stationarity and must be controlled during both data collection and analysis.

\subsubsection{Environmental Noise and Physiological Artifacts}

EEG recordings are highly susceptible to contamination from both internal and external sources. Internally, eye movements (Electrooculography (EOG)), muscle contractions (Electromyography (EMG)), and even jaw clenching introduce high-amplitude, non-stationary artifacts that often overlap with EEG frequency bands of interest~\cite{Mousavi02102019,7319297,app132312800}. Externally, electromagnetic interference from power sources, mobile devices, or lighting systems can further degrade signal quality~\cite{PHILIP2022824}.

\begin{figure}[H]
    \centering
    \includegraphics[width=0.8\textwidth]{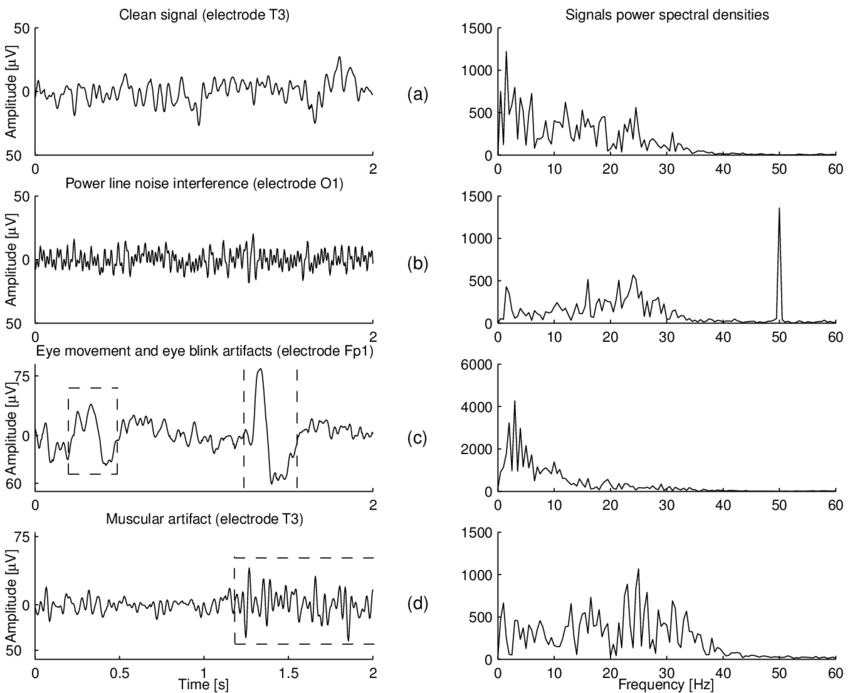}
    \caption{EEG signals perturbed by various artifacts, including power line noise, eye movements, eye blinks, and muscle movements, along with their corresponding power spectral densities (PSDs). These artifacts overlap with frequency bands of interest, complicating signal analysis. Adapted from Nicolas-Alonso and Gomez-Gil (2012) \citep{GarciaMolina2004}.}
    \label{fig:eeg_artifacts}
\end{figure}

Figure~\ref{fig:eeg_artifacts} shows how these artifacts distort the temporal and spectral characteristics of EEG signals, complicating preprocessing and classification pipelines. Their unpredictable nature contributes significantly to temporal drift in BCI performance.

\subsubsection{Learning and Adaptation Effects}

\begin{figure}[H]
    \centering
    \includegraphics[width=0.8\textwidth]{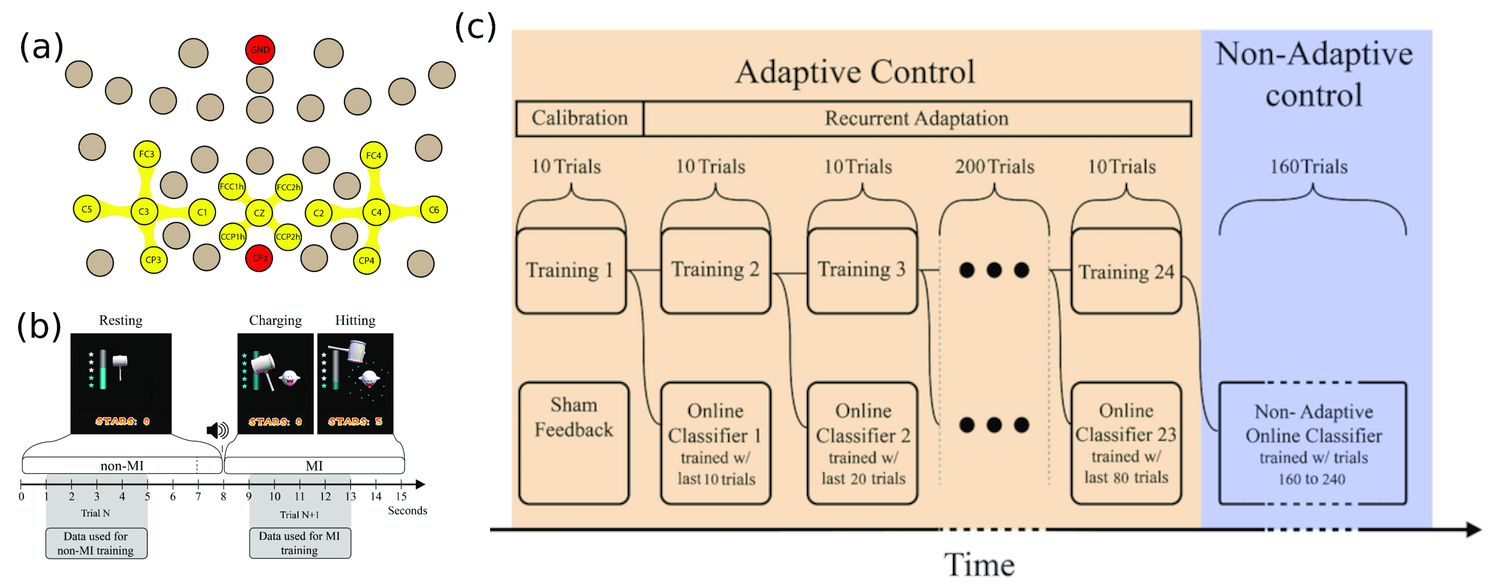}
    \caption{EEG channel configuration over the sensorimotor cortex and the MI training paradigm illustrating learning-induced adaptation in BCI systems. Initially, broad cortical activations become more localized with practice, necessitating the model's adaptive recalibration. Adapted from Rezeika et al. (2021) \citep{Rezeika2021}.}
    \label{fig:learning_adaptation1}
\end{figure}

Another important source of non-stationarity arises from neuroplasticity during BCI use. As users become more proficient at controlling the BCI, their neural activation patterns undergo measurable changes. Initially, widespread cortical areas may be recruited, but over time, activity becomes more focused and consistent~\cite{Mousavi02102019,raza2016adaptive,Samek_2012,10737224,10247260}. These changes in neural strategy reflect co-adaptation between the user and the system, and require models that can adapt accordingly~\cite{KAPGATE2024100109}.

Figures~\ref{fig:learning_adaptation1} depict how repeated training and task feedback alter EEG dynamics. Such learning-induced changes constitute a unique form of non-stationarity that requires dynamic co-adaptive frameworks~\cite{raza2016adaptive}.

\subsubsection{Task-Specific and Contextual Variations}

The nature of the task and its context significantly impact EEG signals. Even when the core task remains constant, changes in feedback modality, visual presentation, or user expectations can lead to altered neural responses and cortical reorganization~\cite{Mousavi02102019, karpowicz2025stabilizing, 10737224,5638613}. Feedback latency, reward schemes, and interface design also influence EEG activity and contribute to variability across sessions~\cite{9195541, 9757884, 10247260}. Such task-dependent variability induces non-stationarities that challenge the generalization of EEG classifiers across sessions and subjects~\cite{ZHANG2021106150, 6945117, 7319297}.

\begin{figure}[H]
    \centering
    \includegraphics[width=0.9\textwidth]{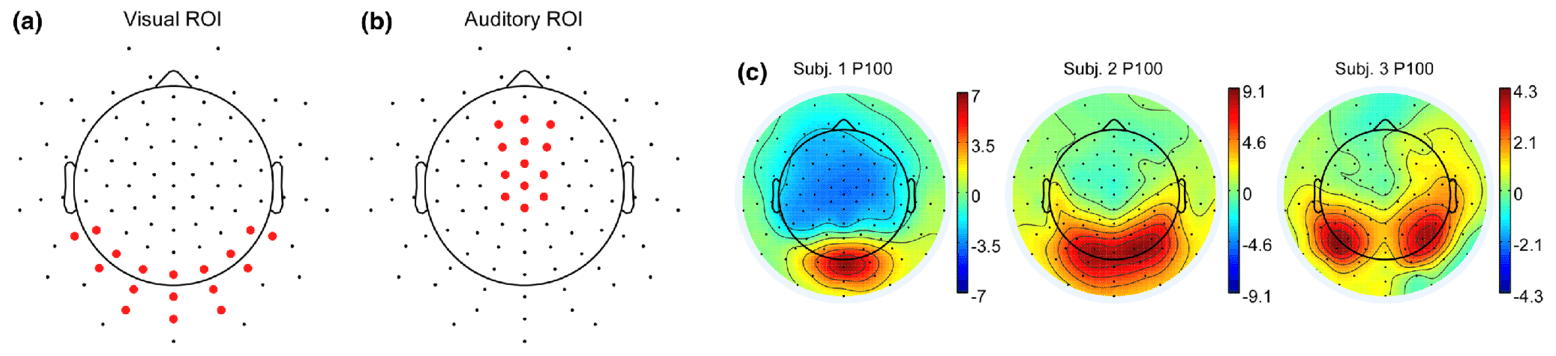}
    \caption{EEG ROI comparison for visual (left) and auditory (right) tasks. Distinct cortical regions are activated depending on sensory modality, highlighting the impact of task context on EEG signal patterns~\cite{Sato2015}.}
    \label{fig:eeg_context}
\end{figure}

Figure~\ref{fig:eeg_context} shows how sensory modality (visual vs. auditory) changes the cortical regions activated during task performance. Such context-dependent variability underscores the importance of standardizing experimental conditions or designing models robust to task diversity.

\subsubsection{Motor Imagery Paradigm}

MI-based BCIs are among the most extensively studied paradigms, relying on imagined limb movements to generate distinguishable EEG patterns. However, MI signals exhibit significant non-stationarity due to factors such as fatigue, attention, and session variability.

Several studies have addressed non-stationarity in motor imagery BCIs through spatial filtering~\cite{Samek_2012,ZHANG2021106150,9757884}, adaptive learning~\cite{8239668,9760113,10737224}, and deep learning~\cite{8897723,10083486}, though generalization across users remains a challenge.

\subsubsection{Covariate Shift and Temporal Drift}

From a machine learning standpoint, EEG non-stationarity can also be interpreted through the lens of covariate shift and temporal drift. \textbf{Covariate shift} refers to a change in the input distribution $P(X)$ over time, even if the underlying relationship $P(Y|X)$ remains unchanged~\cite{YAMADA20102353,8239668,5597639}. This leads to a mismatch between training and deployment conditions, resulting in reduced classifier accuracy.

\textbf{Temporal drift}, whether abrupt or gradual, may be caused by spontaneous attention changes, fatigue, habituation, or external perturbations. Such drifts necessitate adaptive algorithms capable of tracking changes in data distributions to preserve classification performance~\cite{4100850, 5638613, 7319297}. Figure~\ref{fig:physio_fluctuations} visually simulates this drift over time.

To address these challenges, it is possible to reweight the training data, perform domain adaptation~\cite{8852284,6999160}, and use incremental learning techniques that adapt models in real-time to non-stationary data.

\subsection{Implications and Future Directions}

The pervasive nature of EEG signal non-stationarity challenges the conventional assumptions of static machine learning models and limits the scalability of BCI systems. Traditional approaches often require extensive calibration and session-specific tuning, which is impractical in real-world or clinical applications. As BCI technology moves towards real-time deployment, wearable systems, and user-independent interfaces, addressing non-stationarity becomes not just important—but essential.

Recent efforts leveraging transfer learning, domain-invariant representation learning, and online adaptation have shown promise in mitigating these effects~\cite{9154600,6177271,NICOLASALONSO2015186}. Furthermore, hybrid systems integrating multiple modalities, such as SSVEP combined with P300 responses, have demonstrated improved resilience to non-stationarity through adaptive classification schemes~\cite{KAPGATE2024100109}. The design of robust preprocessing pipelines, artifact-aware training paradigms, and co-adaptive learning systems remains an active area of research aimed at bridging the gap between lab-grade BCIs and practical applications.

In summary, EEG signal non-stationarity is a multifaceted phenomenon rooted in biological, technical, and contextual dynamics. Its presence necessitates a paradigm shift in BCI system design—one that embraces adaptability, personalization, and robustness as core principles.

\section{Review methodology}
\label{section:review}

This section outlines the systematic approach used to review the literature on handling non-stationarity in EEG-based BCI systems. The review focused on works that explored the intersection of non-stationary EEG signals and explainable AI (XAI) within the context of BCI. A total of 115 papers were manually analyzed to identify trends, themes, and methodological developments.

\subsection{How the review was conducted, how the papers were selected?}

The methodology employed for conducting the literature review on EEG-based BCI systems, with a focus on non-stationarity and XAI, ensured comprehensive coverage of relevant publications through a transparent selection process and thematic organization of the final corpus.

\subsection{Paper Search and Selection Procedure}

The literature search was conducted systematically using a combination of automated and manual strategies to ensure both breadth and depth:

\begin{enumerate}[nosep]
    \item Databases searched included IEEE Xplore, Elsevier ScienceDirect, SpringerLink, MDPI, Google Scholar, and PubMed to retrieve relevant peer-reviewed articles and preprints.
    \item Predefined keywords and Boolean combinations such as ("EEG" AND "non-stationarity"), ("covariate shift" AND "BCI"), ("adaptive BCI" OR "online learning"), ("transfer learning" AND "EEG"), and ("explainable AI" AND "brain-computer interface") were used.
    \item The review included papers published between 2004 and early 2025, covering two decades of developments in the field.
    \item Inclusion criteria consisted of focus on EEG-based BCI systems, explicit addressing of non-stationarity or covariate shift, presentation of technical, algorithmic, or methodological contributions, and use or discussion of explainable AI techniques (in whole or in part).
    \item Exclusion criteria filtered out studies focused purely on hardware or sensor design, review papers without original methodological contributions, and articles dealing with BCI applications not involving EEG or non-stationarity.
\end{enumerate}

\subsection{Screening and Curation Process}

The initial search returned over 200 articles. A three-stage screening process was applied to refine the selection:
\begin{enumerate}[nosep]
    \item Title and abstract screening filtered out obvious mismatches.
    \item Full-text reviews assessed relevance to non-stationarity, EEG, and methodological depth.
    \item A total of 76 articles meeting all criteria were included in the final corpus and assigned unique IDs for annotation and thematic analysis, while 134 references were cited across these papers, with statistical analysis performed on the subset of 76 articles.
\end{enumerate}

\subsection{Post-Selection Processing: Manual Clustering and Tagging}

After selection, the 115 papers were subjected to manual thematic analysis based on their methodological focus (see Figure~\ref{fig:Clusters}). Each article was labeled according to one dominant approach for addressing non-stationarity, such as adaptive learning, spatial filtering, or domain adaptation. This allowed for the creation of five meaningful clusters representing the primary research strategies, which are discussed in detail in the following sections.

\subsection{Paper Selection Criteria}

The selection process followed a structured methodology to ensure relevance and quality:
\begin{enumerate}[nosep]
    \item Articles were collected from IEEE Xplore, Elsevier ScienceDirect, Springer, MDPI, and Google Scholar, including only peer-reviewed journals, conference proceedings, and preprints with relevant methodological content.
    \item The review considered works published between 2004 and early 2025, covering two decades of research.
    \item Search terms included combinations of “EEG”, “non-stationarity”, “covariate shift”, “BCI”, “adaptive BCI”, “transfer learning”, and “explainable AI”.
    \item Studies explicitly addressing EEG signal non-stationarity, transfer learning, domain adaptation, spatial filtering, or model interpretability in the BCI context were retained.
    \item Articles lacking methodological depth, focused solely on hardware design, or not involving EEG were excluded.
\end{enumerate}

\subsection{Temporal Publication Trends}

\begin{figure}[H]
    \centering
    \includegraphics[width=1\textwidth]{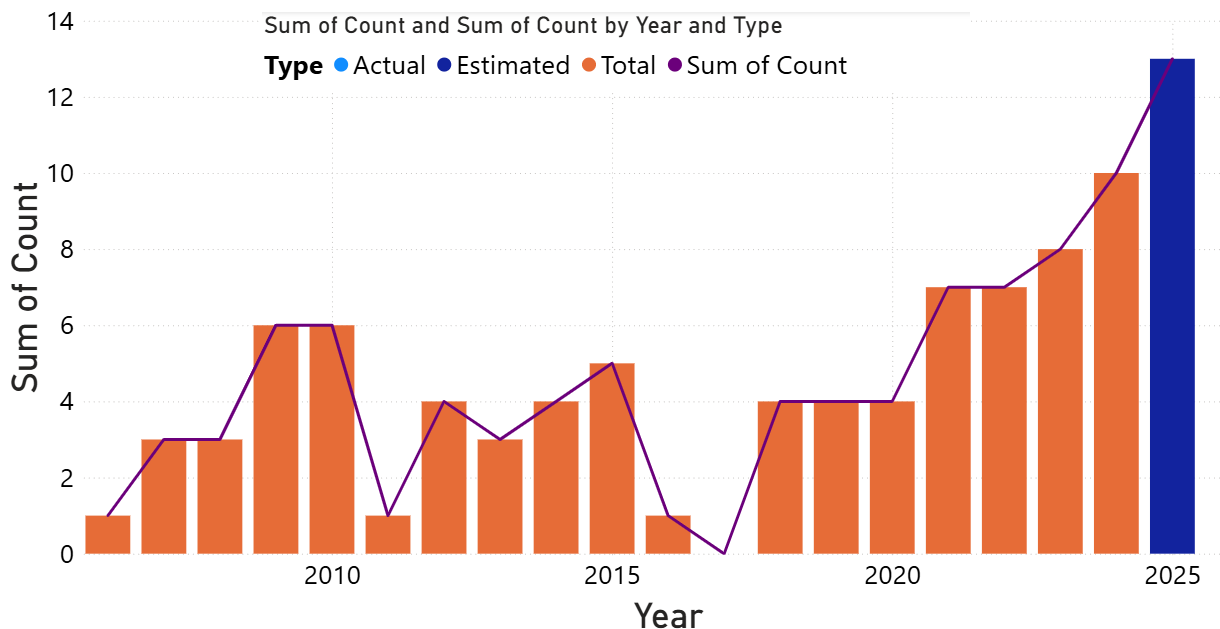} 
    \caption{Papers published over the years within this topic}
    \label{fig:Papers published over the years}
\end{figure}

The bar graph in Fig.~\ref{fig:Papers published over the years} presents the annual count of EEG-based BCI research articles addressing non-stationarity.
    \begin{itemize}[nosep]
        \item From 2004 to 2010, a rising trend marks the foundational phase of research on non-stationarity in EEG-based BCIs, reflecting growing interest and early methodological developments.
        \item The period from 2011 to 2017 saw fluctuations and diversification driven by evolving technical challenges and thematic exploration across subfields.
        \item From 2018 to 2025, substantial growth driven by AI advancements and interdisciplinary efforts was observed, with a dip in 2024–2025 likely due to indexing delays; the 2025 bar indicates projected estimates.
    \end{itemize}

Overall, the figure illustrates a maturing research field that is progressing from foundational studies to sustained expansion, with a focus on practical and adaptive BCI systems.

\subsection{Distribution by Publication Type}

\begin{figure}[H]
    \centering
    \begin{tabular}{@{}cc@{}}
    \includegraphics[width=0.5\textwidth]{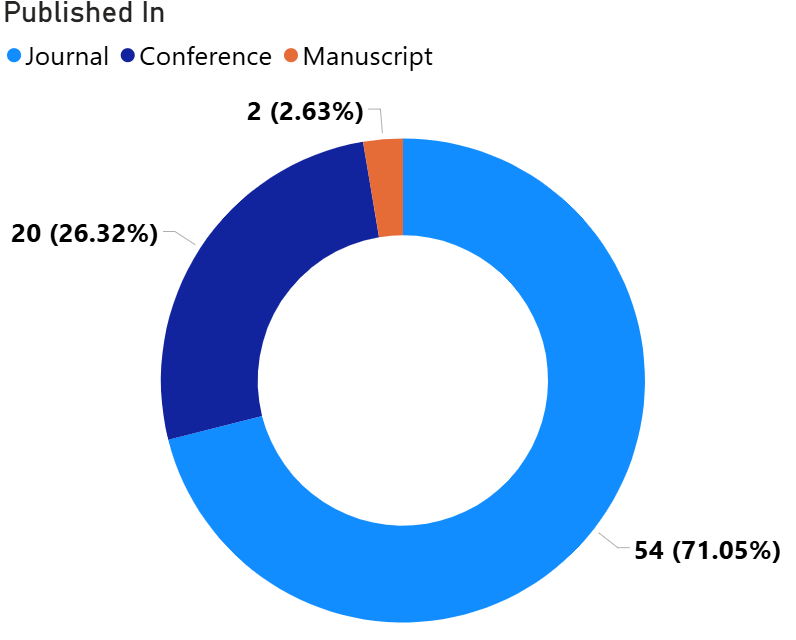} &
     \includegraphics[width=0.5\textwidth]{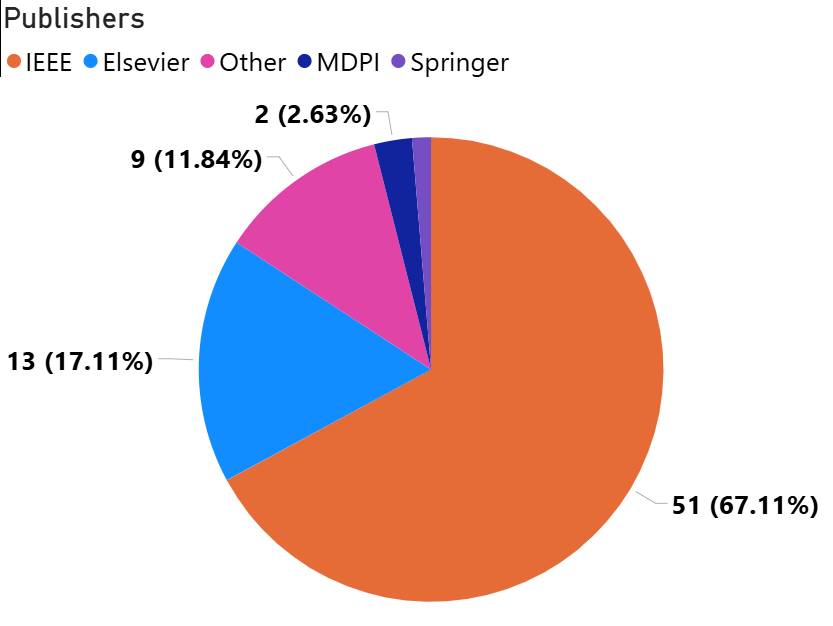}
     \end{tabular}
    \caption{Distribution of Published Papers in Journals vs Conferences (left), distribution of publications by publisher (right).}
    \label{fig:Conference or Journal}
\end{figure}

Fig.~\ref{fig:Conference or Journal} shows the types of venues where the reviewed works were published:
\begin{enumerate}[nosep]
    \item Journals comprised 71.05\% (54 papers), indicating a preference for in-depth, peer-reviewed dissemination.
    \item Conferences accounted for 26.32\% (20 papers), reflecting the field’s dynamic and fast-evolving nature with emphasis on early-stage results.
    \item Manuscripts represented 2.63\% (2 papers), typically preprints, white papers, or unpublished studies.
\end{enumerate}

\subsection{Distribution by Publisher}

%

Figure~\ref{fig:Conference or Journal} (right panel) depicts the distribution of publications across various publishers represented in this review:

\begin{itemize}[nosep]
    \item IEEE led with 67.11\% (51 publications), reflecting its specialization in engineering, signal processing, and neural interface technologies.
    \item Elsevier contributed 17.11\% (13 publications), emphasizing biomedical engineering and neuroscience.
    \item Other publishers made up 11.84\% (9 publications), adding diverse and specialized insights.
    \item MDPI and Springer represented 2.63\% (2 publications) and 1.32\% (1 publication), respectively.
\end{itemize}

This distribution underscores IEEE’s central role while highlighting valuable contributions from other publishers, enriching the multidisciplinary landscape.

\subsection{Purpose of Explainable AI in Non-Stationarity Research}

\begin{figure}[H]
    \centering
    \includegraphics[width=1\linewidth]{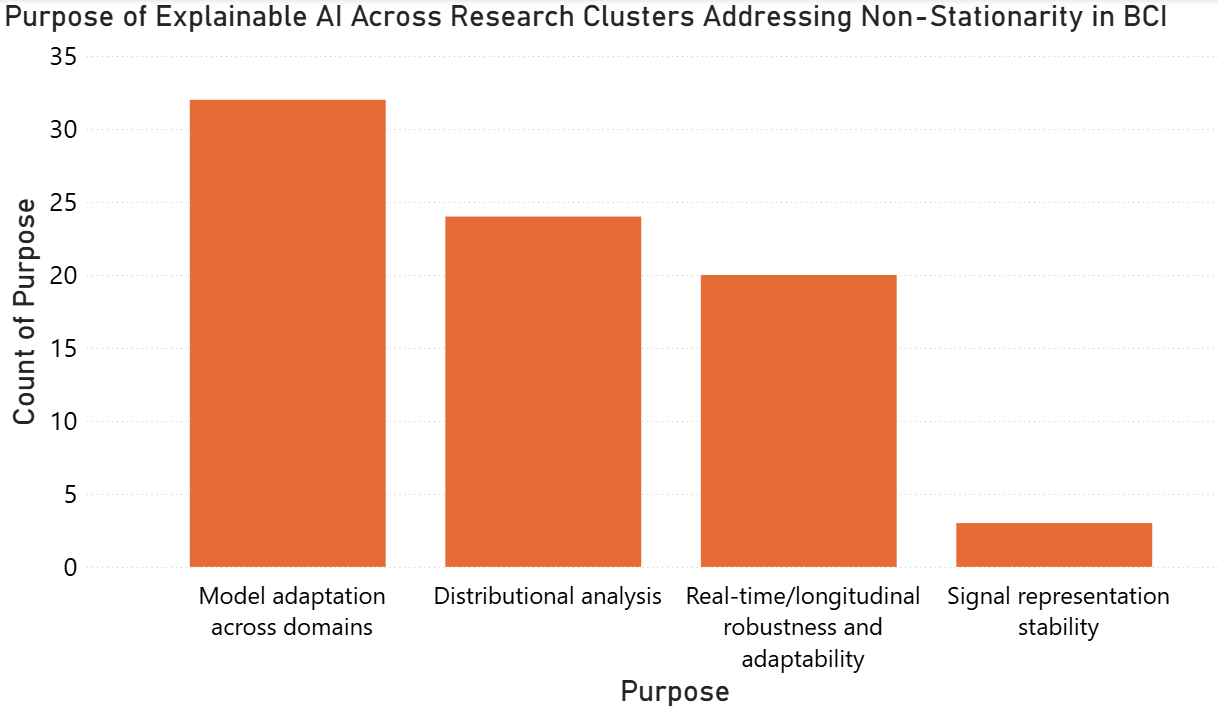}  
    \caption{Distribution of research purposes for applying XAI in BCI studies addressing non-stationarity. The majority of studies focus on real-time and longitudinal robustness, followed by domain adaptation and distributional analysis. Signal representation stability is the least commonly addressed purpose.}
    \label{fig:Purpose of Explainable AI Across Research Clusters Addressing Non-Stationarity in BCI}
\end{figure}

Figure~\ref{fig:Purpose of Explainable AI Across Research Clusters Addressing Non-Stationarity in BCI} outlines the intent behind using XAI methods in non-stationary EEG BCI research:

\begin{enumerate}[nosep]
    \item \textbf{Model Adaptation Across Domains:} \textasciitilde32 studies
    \item \textbf{Distributional Analysis:} \textasciitilde24 studies
    \item \textbf{Real-Time/Longitudinal Robustness and Adaptability:} \textasciitilde20 studies
    \item \textbf{Signal Representation Stability:} \textasciitilde3 studies
\end{enumerate}

The emphasis lies on enhancing adaptability and generalization of BCI systems, while few studies focus on stabilizing underlying signal representations.

\subsection{Clustering of Research Papers}

\begin{figure}[H]
    \centering
    \includegraphics[width=0.5\textwidth]{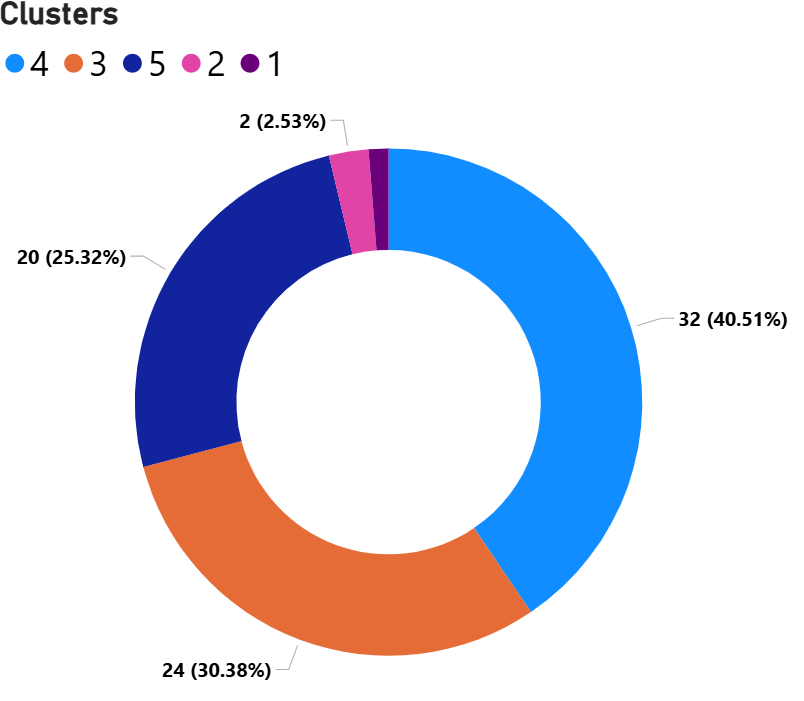}
    \caption{The chart shows the manual thematic clustering of BCI research papers addressing EEG non-stationarity and covariate shift. The distribution of papers across five clusters is illustrated by different colors: Cluster 1 (purple) addresses covariate shift mitigation; Cluster 2 (pink) emphasizes transfer learning and domain adaptation strategies; Cluster 3 (orange) covers Riemannian geometry-based methods; Cluster 4 (dark blue) highlights adaptive and online learning approaches; and Cluster 5 (blue) focuses on spatial filtering and feature extraction techniques. Cluster proportions reflect their relative representation in the literature.}
    \label{fig:Clusters}
\end{figure}

To analyze the methodological diversity, the reviewed papers were manually grouped into five clusters based on thematic similarities (Figure~\ref{fig:Clusters}). Titles, abstracts, and keywords were used to infer the dominant methodological focus. Where papers spanned multiple themes, the primary focus was used for classification.

\begin{enumerate}
    \item \textbf{Cluster 1 (1.27\%, Purple):} Covariate shift and non-stationarity mitigation techniques.
    \item \textbf{Cluster 2 (2.53\%, Pink):} Transfer learning and domain adaptation strategies.
    \item \textbf{Cluster 3 (30.38\%, Orange):} Riemannian geometry-based approaches for manifold learning.
    \item \textbf{Cluster 4 (40.51\%, Blue):} Adaptive and online learning systems for real-time BCI.
    \item \textbf{Cluster 5 (25.32\%, Dark Blue):} Spatial filtering and feature extraction (e.g., CSP, SSA).
\end{enumerate}

These clusters highlight primary research directions that address EEG signal variability and enable robust BCI systems across time, users, and conditions.

\subsection{Manual Clustering Methodology}

To analyze the diverse strategies employed in handling non-stationarity in EEG-based BCI research, a manual thematic clustering of the selected papers was conducted. The goal was to group research articles based on their primary methodological approach or purpose. The process followed a structured and iterative review strategy, detailed below:

\begin{enumerate}
    \item Titles and keywords were reviewed to identify high-level themes such as domain adaptation, covariate shift, Riemannian geometry, adaptive learning, or spatial filtering, allowing preliminary categorization.
    \item Abstracts and methodology sections were examined to clarify main contributions for ambiguous cases, ensuring clustering based on actual techniques rather than superficial indicators.
    \item Five distinct clusters were defined based on thematic content: Covariate shift \& non-stationarity mitigation, Transfer learning \& domain adaptation, Riemannian geometry-based methods, Adaptive BCI systems \& online learning, and Spatial filtering \& feature extraction.
    \item Each paper was assigned to one cluster based on dominant methodology, with priority given to the main contribution when multiple clusters applied.
    \item Initial assignments were independently verified and refined for thematic consistency, minimizing subjective bias and ensuring precise categorization.
    \item A quantitative summary was generated by tallying papers per cluster, visualized in Figure~\ref{fig:Clusters}.
\end{enumerate}

This manual clustering method enabled a nuanced grouping of papers based on their technical focus, providing a clear understanding of the dominant trends and methodological diversity in the literature.

\section{Approaches and Solutions to EEG Non-Stationarity and Domain Adaptation in BCIs}
\label{section: approaches}

EEG-based BCIs face significant challenges due to \textit{non-stationarity} — temporal changes in statistical properties of neural signals — and \textit{covariate shift}, where differences between training and testing feature distributions arise from inter-session variability, inter-subject differences, or environmental factors. Such variations stem from physiological fluctuations, changes in mental states, electrode repositioning, hardware inconsistencies, and artifacts, which together cause notable performance degradation in laboratory and real-world BCI applications~\cite{app132312800, Samek_2012, 6482603, 8701679, 9134411, doi:10.1073/pnas.1221127110, Chase2012visuomotor}.

To mitigate these issues, research has focused primarily on two complementary classes of solutions: first, signal normalization and feature space alignment strategies that stabilize signal distributions and reduce domain mismatches before classification; and second, classifier adaptation algorithms that dynamically adjust or transfer models to maintain robust performance under distributional shifts~\cite{9076296, 9782441, 9034072, 4803844, 8337789, 5305646, APICELLA2023106205}.

Signal normalization methods operate at the raw signal or feature level and include classical preprocessing steps such as high-pass filtering, which removes slow baseline drifts and low-frequency artifacts below 0.5–1 Hz to improve data stationarity and feature robustness~\cite{hjorth1975}. Detrending techniques similarly address slow linear or nonlinear trends caused by electrode impedance changes or subject movement, stabilizing the mean baseline over time through polynomial fitting or adaptive filtering~\cite{hjorth1975}. Surface Laplacian transformations, also known as current source density methods, compute the second spatial derivative of scalp potentials to emphasize local cortical activity while attenuating volume conduction and reference electrode effects, thereby enhancing spatial resolution and signal-to-noise ratio, especially for motor imagery and event-related potential paradigms~\cite{hjorth1975,kayser2015,mcfarland2008,tenke2012}.

Building on spatial filtering, several approaches constrain the classical Common Spatial Patterns (CSP) framework to improve feature stationarity across trials or sessions. For instance, the stationary CSP (sCSP) algorithm maximizes class separability while penalizing filters sensitive to inter-trial covariance variability~\cite{Samek_2012, 6482603, 9757884}, and its Tikhonov-regularized variant (sTRCSP) reduces overfitting and noise sensitivity~\cite{Samek_2012, 6782557}. Similarly, shared subspace CSP (ssCSP) identifies non-stationary components common across subjects through difference covariance analysis and PCA, incorporating these as regularizers during CSP computation~\cite{Samek_2012, 4785192}. Relatedly, subspace analysis techniques such as group SSA enforce intra-class stationarity by minimizing divergence measures between class-specific groups, while discriminative SSA balances stationarity with preservation of class discriminability~\cite{6482603, 9757884, 6782557}. These methods aim to retain task-relevant components that might otherwise be lost through aggressive artifact removal~\cite{4682649}.

Frequency-selective and target-aligned filtering further enhances cross-subject generalization by decomposing signals into sub-bands (e.g., 8–32 Hz) and aligning source data covariance matrices into the target domain using Euclidean alignment techniques. Feature selection via minimum redundancy maximum relevance (mRMR) combined with linear discriminant analysis (LDA) classifiers then yields robust motor imagery classification~\cite{ZHANG2021106150, 10684016}. Covariance matrix alignment is also pursued within both Euclidean and Riemannian geometry frameworks. Euclidean Alignment (EA) normalizes covariance matrices to the identity matrix, improving cross-subject comparability, and has been extended to cross-species adaptation (e.g., mouse-to-human EEG) \cite{10684016, 9905704}. Riemannian Alignment (RA) leverages the geometry of symmetric positive definite matrices by mapping covariances to a common reference, typically the Riemannian mean, preserving manifold structure and enhancing classifiers such as Minimum Distance to Riemannian Mean (MDRM) and Tangent Space LDA (TSLDA)~\cite{6609172, 8013808}. Source-free domain adaptation methods like AEA further refine EA by learning lightweight projections aligning target data to a frozen model’s feature space without requiring source data access, which is critical for privacy-preserving BCI deployments \cite{10684016}.

Robustness is improved through channel and trial selection strategies, which identify subsets of electrodes and data samples less affected by non-stationarity. Sequential Floating Backward Selection (SFBS) selects channels with stable signal characteristics \cite{8512974, s21093225}, while trial selection methods pick auxiliary trials closest to the target domain in Riemannian space, reducing training–testing mismatch \cite{10423987}. Additionally, abrupt shifts in covariance structures are detected using graph-based stationarity measures such as STATIO, which exploits wide-sense graph stationarity properties to reduce computational complexity, and other methods, including exponentially weighted moving average (EWMA) statistics or Isolation Forest anomaly detection, enabling timely correction of covariate shifts \cite{app132312800, 4803844}.

Classifier adaptation techniques dynamically update decision boundaries or model parameters to cope with evolving EEG distributions. Incremental algorithms, such as Sequential Updating Semi-supervised Spectral Regression KDA (SUSS-SRKDA), incorporate new unlabeled data through efficient matrix updates and session-wise feature mean normalization, facilitating continual learning~\cite{4100850}. In clinical contexts, adaptive support vector machines incrementally update support vectors across sessions to improve long-term performance, as demonstrated in stroke rehabilitation \cite{4803844}. Hybrid BCIs combining SSVEP and P300 paradigms benefit from continuous LDA parameter updates to maintain classification resilience amid gradual non-stationarity~\cite{10035017}. Real-time decoding for motor control tasks is enhanced by adaptive probabilistic neural networks that update model parameters online to improve accuracy during dynamic hand grasp control~\cite{HAZRATI2010730}. Moreover, bilateral adaptation frameworks simultaneously adjust both user strategies and system parameters via neurofeedback, accelerating convergence and promoting stable co-adaptive learning in closed-loop BCIs \cite{LI2010373}.

Transfer learning and domain adaptation frameworks have gained prominence to leverage data from related domains or subjects. Improved Gaussian mixture models with enhanced parameter initialization accelerate unsupervised adaptation in dynamic BCI environments \cite{Liu2010iGMM}. Prototypical Contrastive Domain Adaptation (PCDA) combines source selection, Euclidean Alignment of covariance matrices, adversarial domain alignment, and prototypical contrastive loss to learn domain-invariant representations for motor imagery classification~\cite{9076296}. Selective Multi-source Domain Adaptation (Selective-MDA) dynamically weights multiple source domains based on discriminator loss, improving robustness when source-target similarities vary~\cite{8337789}. Electrode-level Domain Adaptation Networks (EDAN) explicitly model variability at electrode granularity via spatial–temporal convolutional neural networks with attention, tackling both intra- and inter-subject differences~\cite{10423987}. In intracortical BCIs, NoMAD aligns neural activity to stable latent manifolds learned through recurrent neural networks, compensating for neural signal degradation and electrode instability without extensive recalibration~\cite{9354668}. Domain adaptation techniques have also been successfully applied to SSVEP classification, reducing calibration efforts across sessions in both online and offline contexts~\cite{10098737, 7580614}.

Recent deep learning approaches integrate domain adaptation directly into the representation learning process. Multiview Adversarial Contrastive Networks (MACNet) extract complementary Euclidean and Riemannian features, applying adversarial and contrastive learning alongside domain mix-up augmentation to achieve robust cross-subject generalization with limited labeled data~\cite{10436703}. Autoencoder-based convolutional neural networks (AE-CNNs) learn invariant latent representations by reconstructing EEG signals and exploiting residuals to isolate session-specific variations, facilitating adaptation \cite{9034072}.

Data augmentation and simulation-based strategies further enhance adaptability. Unsupervised self-paced learning methods enable classifier updates through user interaction without explicit labels, making them particularly useful in game-based BCI systems \cite{HASAN2012598}. Generative adversarial networks conditioned on task variables synthesize realistic EEG trials to augment training data for rare or underrepresented classes, improving classifier robustness under varying attention levels \cite{app132312800, 9134411}. Simulation studies modeling neural signal degradation, such as firing rate declines, provide testbeds for retraining and evaluating adaptive decoders.

Collectively, these methods demonstrate a broad spectrum of solutions addressing EEG non-stationarity, ranging from preprocessing and normalization approaches, such as sCSP, RA, EA, and STATIO, to advanced adaptation frameworks, including PCDA, MACNet, and NoMAD. Current trends emphasize combining spatial, spectral, and geometric normalization with adaptive classification~\cite{9076296, 9905704, 5305646}; exploiting unlabeled target data for unsupervised or source-free adaptation \cite{10684016, Zhao_Yan_Lu_2021, HASAN2012598}; modeling variability at multiple scales, from individual electrodes to full covariance structures~\cite{10423987}; incorporating physiologically inspired constraints to enhance interpretability~\cite{8337789, 4540104, 4682649}; employing generative models and simulations to anticipate rare or future distribution shifts~\cite{app132312800, Zhao_Yan_Lu_2021}; and exploring multiple time scales of closed-loop adaptation for improved co-adaptive learning \cite{6091387}.

Several open-source toolboxes enable the practical implementation of non-stationarity mitigation and domain adaptation techniques in both Python and MATLAB. 
In Python, MNE-Python provides built-in CSP workflows via \texttt{mne.decoding.CSP} for spatial filtering and regularization~\cite{mnePython}. 
The osl-ephys library extends MNE with modular pipelines for preprocessing and source-space analysis~\cite{oslephys2024}, while \textbf{Geomstats} supports Riemannian operations on covariance matrices, facilitating RA implementations~\cite{geomstats2020}. 
For source-connectivity and single-trial feature extraction, the SCoT toolbox offers CSPVARICA and MVARICA algorithms~\cite{scot2014}, and the Gumpy framework enables hybrid EEG/EMG decoding with both traditional and deep learning classifiers~\cite{gumpy2018}. 

In MATLAB, FieldTrip is a widely used platform for advanced spatial filtering, preprocessing, and time-frequency analysis~\cite{fieldtrip}, while EEGLAB provides an extensible environment for ICA-based artifact removal and statistical analysis~\cite{eeglab}. 
The rASRMatlab plugin for EEGLAB implements Riemannian Artifact Subspace Reconstruction (rASR) for robust artifact correction~\cite{rasrMatlab}, and the Neurophysiological Biomarker Toolbox (NBT) offers pipelines for biomarker extraction from EEG/MEG data~\cite{nbt2015}. 
The availability of these toolboxes reduces the barrier to translating advanced non-stationarity handling methods into real-world BCI systems, and they provide reproducible reference implementations for future research. 
Future research directions are likely to focus on domain generalization techniques capable of learning invariant features robust to unseen shifts, multimodal adaptation that integrates EEG with complementary biosignals, and hardware–software co-design strategies aimed at enabling long-term, calibration-free BCI operation.

\section{Discussion}
\label{section:discussion}

\subsection{Key Methods for Dealing with EEG Non-Stationarity}

The literature review indicated that non-stationarity is a critical issue in EEG-based BCI systems, as the statistical properties of EEG signals can vary significantly over time, across sessions, and between subjects~\cite{Samek_2012, 8897723}. This variability poses a challenge for building robust classifiers, which often assume stationarity in the training and test distributions. Consequently, a wide range of methods has been developed to detect and address these distributional shifts.

There are multiple origins of non-stationarity in the EEG signal, and these causes can lead to changes that may be misinterpreted. For example, gel drying over time can reduce signal quality, potentially leading to incorrect interpretations of EEG signals when measuring brain responses related to fatigue or prolonged cognitive tasks. The brain-evoked responses may change not because of fatigue or other mental processes, but because of a decrease in signal quality over time. Therefore, experimental paradigms should consider such effects and mitigate them through appropriate signal processing techniques to normalize the signals. Any progressive change over time during an experiment can originate from a decrease in signal quality, rather than changes in the evoked brain responses. In BCI applications, identifying the source of change helps determine whether a patient needs rest, the gel has dried, or other causes are present.

Primary algorithms for detecting non-stationarity or covariate shifts include statistical monitoring techniques such as EWMA ~\cite{app132312800, 5638613}, kernel-based measures such as Maximum Mean Discrepancy (MMD) ~\cite{YAMADA20102353, 10247260}, and signal-space transformations such as Principal Component Analysis (PCA) and Independent Component Analysis (ICA) ~\cite{7319297, 8512974, 1639190}. These methods can identify changes in data distributions that may impact classification accuracy and are typically integrated into preprocessing pipelines or adaptation strategies~\cite{4100850, 6609172}.

Addressing EEG non-stationarity begins at the raw signal stage rather than at the classification level. Preprocessing steps such as low-pass filtering, high-pass filtering, and detrending are essential to account for slow drifts, motion artifacts, and baseline shifts that could introduce non-stationary patterns in the signal~\cite{Mousavi02102019, 8512974, 10083486, 6609172}. Effective preprocessing can stabilize the signal and help improve the robustness of downstream models~\cite{Samek_2012, 6782557}.

\subsection{Strategies to Address Covariate Shift}

In dynamically changing environments, three main strategies have emerged to address covariate shifts in EEG-based BCI systems:

\begin{enumerate}
    \item \textbf{Invariant Feature Extraction:} This approach involves limiting the feature space to components that remain stable across sessions or subjects. Methods such as Stationary Common Spatial Patterns (sCSP)~\cite{Samek_2012}, robust feature construction~\cite{6782557}, and sub-band filtering with transfer learning~\cite{ZHANG2021106150} aim to extract features that are less sensitive to non-stationarity, thereby improving cross-session and cross-subject generalization.

    \item \textbf{Reference-Based Normalization:} This method constructs a reference distribution or model from the training data and aligns new incoming data to this reference, effectively reducing domain mismatch. Target alignment techniques using Riemannian geometry~\cite{8013808}, sub-band target alignment CSP~\cite{ZHANG2021106150}, and covariance matrix alignment~\cite{6482603} exemplify this strategy, facilitating transfer learning across subjects or sessions. These approaches use strategies similar to whitening the signal to minimize the changes between sessions. Target alignment can be performed when the data is streamed, trial by trial, through batch mode with sequences of trials, or for the entire session.

    \item \textbf{Adaptive Learning:} Adaptive classifiers update their parameters continuously or semi-supervisedly to track evolving signal properties. Online recursive ICA methods~\cite{7319297}, unsupervised adaptation of LDA~\cite{5638613}, and semi-supervised domain adaptation networks~\cite{10247260} have demonstrated effectiveness in maintaining classification performance in the presence of gradual or abrupt covariate shifts.
\end{enumerate}

\subsection{Limitations}

Despite notable progress in algorithmic development, several limitations continue to constrain the comprehensive evaluation of covariate shift mitigation techniques in BCI research. A key challenge is that widely used datasets—particularly those from earlier BCI competitions—often include data from a small number of participants (e.g., nine), with limited sessions and few trials per class ~\cite{6945117, ZHANG2021106150}. While such datasets have been used for assessing covariate shifts, it is necessary to separate outliers from long covariate shifts.

More recent datasets have increased participant numbers but often lack sufficient session-level variability, which is essential for evaluating session-to-session non-stationarity handling methods ~\cite{app132312800}. The scarcity of datasets with multiple sessions per subject limits the ability to benchmark adaptive learning and covariate shift detection techniques effectively~\cite{10083486, 7319297}. More information should also be provided about the sessions, including their dates and times. These limitations highlight key areas where the field must evolve—not only in data collection and experimental design, but also in algorithmic expectations and evaluation standards.

\subsection{Future Directions}

BCI research has focused on decoding brain activity using advanced machine learning and signal processing methods. However, neural engineering alone cannot make BCIs reliable~\cite{Wolpaw_2025}. BCI is also a skill that must be developed over multiple sessions, with performance in attention- or motor imagery-based tasks improving through practice. To address the limitations discussed above and advance the field holistically, we propose the following future directions, organized under three major themes:

\subsubsection{Data and Evaluation Frameworks}

Future work should prioritize the development and open dissemination of benchmark datasets with multiple well-annotated sessions per subject, balanced trial counts, and metadata describing environmental and physiological conditions. Such datasets would provide the empirical foundation for testing non-stationarity handling methods under realistic, ecologically valid scenarios. Standardized protocols for evaluating covariate shift mitigation—across sessions, subjects, and devices—are also needed to enable transparent and reproducible method comparisons.

\subsubsection{Algorithmic Innovation}

Efforts should shift from isolated techniques to hybrid frameworks that integrate invariant feature extraction, signal-space alignment, and adaptive classifiers. Generative modeling approaches, such as VAEs and GANs, can be employed to disentangle task-relevant components from noise or session-specific variability, enabling synthetic augmentation or the creation of domain-invariant features. Temporal modeling methods, including recurrent networks and attention-based drift detection mechanisms, should be explored to track longitudinal changes in neural data.

\subsubsection{User-Centered and Practical Adaptation}

BCI systems should not only adapt to non-stationarity passively, but also incorporate user-centered co-adaptation strategies. These may include dynamic task modulation, personalized feedback, or active learning approaches where the system interacts with the user based on classifier uncertainty or performance decay. Hardware-aware algorithms, robust to sensor variability, motion artifacts, and dry electrode noise, are essential for mobile and real-world operation.
Combining shift detection with explainability frameworks would enable understanding of not just when performance declines, but also why, allowing for targeted interventions—whether algorithmic, procedural, or hardware-based.
Addressing EEG non-stationarity is not just a signal processing problem—it is a systems-level challenge that spans data quality, algorithm design, user adaptation, and hardware robustness. Interdisciplinary solutions and real-world testing are required to close the gap between laboratory and long-term practical BCI performance.

\section{Conclusion}
\label{section:conclusion}

EEG-based BCIs hold immense promise for enabling seamless interaction between the human brain and external systems. However, a central challenge hindering reliable and widespread deployment is the pervasive issue of non-stationarity—variability in EEG signals over time, across sessions, and between users. This review systematically examined the origins of EEG non-stationarity and categorized current solutions into five core strategies: signal normalization, feature alignment, classifier adaptation, domain adaptation, and hybrid approaches. These strategies, while diverse in implementation, collectively aim to mitigate the effects of covariate shifts that degrade BCI performance over time or in new contexts.

Addressing non-stationarity is not solely a classification challenge, but a multi-stage process that begins with raw signal acquisition. From preprocessing techniques that stabilize signal characteristics to advanced domain adaptation methods that align distributions across users or sessions, robust BCI performance requires an integrated pipeline approach. Feature-level strategies such as Stationary CSP and Riemannian alignment show promise for extracting invariant representations, while adaptive classifiers and semi-supervised learning frameworks maintain continuous performance tracking and updating.

Significant limitations remain, as many widely used datasets lack sufficient session diversity, participant variability, and environmental metadata, making it challenging to evaluate generalizability in realistic settings. Current methods are often assessed in isolation, without standardized evaluation protocols or holistic performance benchmarks that reflect the dynamic nature of real-world BCI use.

Future directions include the creation of ecologically valid benchmark datasets with detailed annotations, the development of hybrid models that combine invariant features, adaptive learning, and generative modeling, and the implementation of user-centered adaptation strategies that incorporate explainability, uncertainty-driven feedback, and hardware-aware robustness. Together, these advances can bridge the gap between lab-based performance during short sessions and real-world applicability.

In conclusion, overcoming EEG non-stationarity is a systems-level challenge that requires interdisciplinary collaboration across neuroscience, signal processing, machine learning, human-computer interaction, and hardware design. Addressing the full complexity of this problem—across signal, algorithm, user, and environment—will be key to achieving reliable, adaptive, and accessible BCIs.

\appendix
\setcounter{section}{0}
\renewcommand{\thesection}{\Alph{section}}

\section{List of Acronyms}
\addcontentsline{toc}{section}{List of Acronyms}

\begin{tabular}{ll}
\textbf{Acronym} & \textbf{Full Form} \\
\hline
BCI   & Brain-Computer Interface \\
EEG   & Electroencephalography \\
CSP   & Common Spatial Pattern \\
SSVEP & Steady-State Visual Evoked Potential \\
MI    & Motor Imagery \\
SCSP  & Stationary Common Spatial Pattern \\
LDA   & Linear Discriminant Analysis \\
EA    & Euclidean Alignment \\
ERP   & Event-Related Potential \\
RSVP  & Rapid Serial Visual Presentation \\
PSD   & Power spectral densities \\
EOG   & Electrooculography (EOG) \\
EMG   & Electromyography (EMG) \\
XAI   & Explainable Artificial Intelligence \\
ssCSP & subspace CSP \\
mRMR  & minimum redundancy maximum relevance \\
RA    & Riemannian Alignment \\
MDRM  & Minimum Distance to Riemannian Mean \\
TSLDA & Tangent Space LDA \\
SFBS  & Sequential Floating Backward Selection \\
EWMA  & exponentially weighted moving average \\
SUSS-SRKDA & Sequential Updating Semi-supervised Spectral Regression KDA \\
PCDA  & Prototypical Contrastive Domain Adaptation \\
MDA   & Multi-source Domain Adaptation \\
EDAN  & Electrode-level Domain Adaptation Networks \\
MACNet & Multiview Adversarial Contrastive Networks \\
AE-CNNs & Autoencoder-based convolutional neural networks \\
rASR  & Riemannian Artifact Subspace Reconstruction \\
NBT   & Neurophysiological Biomarker Toolbox \\
MMD   & Maximum Mean Discrepancy \\
PCA   & Principal Component Analysis \\
ICA   & Independent Component Analysis \\
sCSP  & Stationary Common Spatial Patterns \\
TBI   & Traumatic Brain Injury \\
\end{tabular}

\vspace{1em}

\appendix
\section{Keywords and Corresponding References}
\begin{table}[htbp]
\centering
\label{tab:keywords_refs}
\begin{tabular}{p{0.35\linewidth} p{0.55\linewidth}}
\hline
\textbf{Keyword} & \textbf{References} \\
\hline
Covariate shift & \cite{8914414, 6854726, 8978471, 6488742, vishwanath2023reducingintraspeciesinterspeciescovariate, YAMADA20102353, app132312800, 6609172, RAZA2015659, 5415628, sugiyama2007covariate, 8239668, 5597639, 6999160, 7280742, raza2016adaptive} \\
Non-stationarity & \cite{Shen2019_challenge_aBCI, 4811472, PHILIP2022824, 9757884, 6782557, 1639190, s21093225, 8512974, Wan2023_nonstationarity_spikes, RAZA2015659, 7319297, 5597639, 6999160, 7280742, 8852284, KAPGATE2024100109, 9441076, 10083486, NICOLASALONSO2015186} \\
Adaptive & \cite{5305646, 4352845, ROY2022105347, HAZRATI2010730, Chase2012visuomotor, LI2010373, 4540104, doi:10.1073/pnas.1221127110, 6091387, HASAN2012598, Liu2010iGMM, 4682649, 4100850, 5638613, 4785192, 5415628, sugiyama2007covariate, 8239668, 7319297, 5597639, 6999160, 7280742, Farshchian2018Adversarial, 9516951, 9076296, DONG2023104453, 9154600, 9034072, 4803844, 8337789, 10098737, 10684016, WANG2024106742, 10436703, APICELLA2023106205, 7580614, Zhao_Yan_Lu_2021, 10737224, 9195541, 8852284, KAPGATE2024100109, 6177271, 6945117, raza2016adaptive, NICOLASALONSO2015186, 9782441, 9354668, 9905704, 10423987, 9534286, 10035017, 9760113, 10247260} \\
Domain Adaptation & \cite{Farshchian2018Adversarial, 9516951, 9076296, DONG2023104453, 9154600, 9034072, 4803844, 8337789, 10098737, 10684016, WANG2024106742, 10436703, APICELLA2023106205, 7580614, Zhao_Yan_Lu_2021, 10737224, 9195541, 8852284, 9782441, 9354668, 9905704, 10423987, 9534286, 10035017, 9760113, 10247260} \\
P300 & \cite{6401177, KAPGATE2024100109} \\
Motor Imagery & \cite{INCE2009236, 8695803, 9441076, 10083486, 6177271, 6945117, raza2016adaptive, NICOLASALONSO2015186, 9782441, 9354668, 9905704, 10423987, 9534286, 10035017, 9760113, 10247260} \\

\hline
\end{tabular}
\end{table}

\subsection*{Acknowledgment}

This work has been supported by the NIH-R15 NS118581 grant.

\bibliographystyle{unsrtnat}
\bibliography{references}

\end{document}